\documentclass[11pt]{article}
\usepackage{amsmath,amssymb,color,graphics,epsfig,cite}

\usepackage{amsfonts}

\newcommand{\be}{\begin{equation}}
\newcommand{\ee}{\end{equation}}
\newcommand{\bea}{\setlength\arraycolsep{2pt} \begin{eqnarray}}
\newcommand{\eea}{\end{eqnarray}}
\newcommand{\nn}{\nonumber}

\def\ft#1#2{{\textstyle{\frac{\scriptstyle #1}{\scriptstyle #2} } }}
\def\fft#1#2{{\frac{#1}{#2}}}

\def\0{{\sst{(0)}}}
\def\1{{\sst{(1)}}}
\def\2{{\sst{(2)}}}
\def\3{{\sst{(3)}}}
\def\4{{\sst{(4)}}}
\def\5{{\sst{(5)}}}
\def\6{{\sst{(6)}}}
\def\7{{\sst{(7)}}}
\def\8{{\sst{(8)}}}
\def\sst#1{{\scriptscriptstyle #1}}
\def\oneone{\rlap 1\mkern4mu{\rm l}}

\begin{document}

\begin{center}
{\Large {\bf Demagnetizing KBR and New Ricci-flat Rotating Metric}}

\vspace{20pt}

Liang Ma and H. L\"u
		
\vspace{10pt}

{\it Center for Joint Quantum Studies, Department of Physics,\\
School of Science, Tianjin University, Tianjin 300350, China }

\vspace{40pt}

\underline{ABSTRACT}
\end{center}

\vspace{10pt}

We construct a new Ricci-flat metric by demagnetizing the recently reported Kerr-Bertotti-Robinson (KBR) solution. The metric is a deformation of the Kerr metric characterized by a parameter $B$, so that the asymptotic Kerr becomes a regular dome of spindle shape with north and south poles. Despite lacking an asymptotically-flat region, we find that the first law of black hole thermodynamics can be established. Some thermodynamic relations are identical to those of the Kerr black hole, as if the constant $B$ is absent. Our Ricci-flat rotating metric serves a neutral seed for a variety of inequivalent schemes of magnetizing the Schwarzschild and Kerr black holes.

\vfill {\footnotesize maliang0@tju.edu.cn\ \ \ mrhonglu@gmail.com}

\thispagestyle{empty}
\pagebreak



\section{Introduction}

Since Einstein formulated General Relativity in 1916, there have been continuing efforts to construct exact solutions to the Einstein field equations \cite{Stephani:2003tm, Griffiths:2009dfa}. Among these, the four-dimensional Schwarzschild and Kerr black holes are the most important, both theoretically and phenomenologically. With the rapid progress in black hole observations through gravitational waves \cite{LIGOScientific:2016aoc} and shadow images \cite{EventHorizonTelescope:2019dse}, the construction of new exact solutions may have direct applications in astrophysical observations.

One particularly important class of solutions in Einstein-Maxwell (EM) gravity is known as the Melvin universe, including the Schwarzschild-Melvin, Kerr-Melvin and Kerr-Newman-Melvin black holes \cite{Bonnor:1954tis,Melvin:1963qx,Ernst:1976mzr,Ernst:1976bsr}. They describe the corresponding black holes immersed in an external uniform magnetic field. These spacetimes are typically of Petrov type I, but recently a new type-D KBR solution describing a Kerr black hole immersed in an external magnetic field was constructed by Podolsk{\'y} and Ovcharenko \cite{Podolsky:2025tle,Ovcharenko:2025cpm}, indicating the existence of multiple independent magnetization procedures in EM gravity. This work inspired several related new follow-up constructions focusing on the static configurations \cite{Astorino:2025lih,Astorino:2026okd,Ovcharenko:2026byw,Barrientos:2026shy}.

The existence of exact solutions describing the Melvin universe is closely related to some special properties of EM gravity, which admits an supersymmetric extension with enhanced global symmetry \cite{Cremmer:1997ct}. Specifically, the Kaluza-Klein (KK) reduction of EM gravity gives rise to an $SU(2,1)/(SU(2)\times U(1))$ scalar coset with enhanced global $SU(2,1)$ symmetry \cite{Breitenlohner:1987dg,Cremmer:1999du}. A standard transformation of magnetizing black holes leading to the Melvin universe were developed in \cite{Gibbons:2013yq}. Conversely, the Schwarzschild or Kerr black holes can be obtained from the Melvin solutions by the demagnetization transformation, rather than by simply setting the external magnetic field to zero.

In this paper, we apply this demagnetization to the aforementioned KBR solution, rather than just setting its magnetic field $B$ to zero. We obtain a new Ricci-flat rotating metric, with parameter $B$ remaining in the metric. This implies that the interpretation of $B$ is not that of an external magnetic field, but is intrinsically geometrical. The resulting metric provides a deformation of the Kerr solution.

The paper is organized as follows. In Section 2, we construct the new Ricci-flat metric and analyse its global structure. For suitable parameters, the metric describes a black hole and we derive the first law of black hole thermodynamics in Section 3. In Section 4, we consider re-magnetizing the Ricci-flat metric and show that it can be used as a seed for constructing a variety of inequivalent magnetizations of the Schwarzschild and Kerr black holes. We conclude the paper in Section 5. The $SU(2,1)$ global symmetry underlying the (de)magnetization was established in \cite{Gibbons:2013yq}, and we present it in the appendix for self-containment.

\section{New Ricci-flat rotating metric}

\subsection{Local metric}

We obtain our Ricci-flat metric via a demagnetization transformation of the recently constructed KBR black hole \cite{Podolsky:2025tle,Ovcharenko:2025cpm}. The (de)magnetizing transformation in EM gravity was developed in \cite{Gibbons:2013yq} and are reviewed in the Appendix. The metric takes the form
\bea
ds^2&=&-\frac{Q}{\mathbb{L} \Omega ^4}\Big[
\frac{a x^2 }{P_0}d\phi+\frac{dt }{r^4
   (1-x^2) P-a^2 x^4Q}\big(a^2 x^2 \mathbb{M} +\mathbb{L} r^2 \Omega ^2 (1-x^2) P\big)
\Big]^2\cr
&&+\frac{(1-x^2) P}{\Omega ^4 \mathbb{L}}\Big[\frac{r^2 d\phi}{P_0}+\frac{a dt }{r^4 (1-x^2) P-a^2 x^4 Q}\big(r^2 \mathbb{M}+\mathbb{L} Q \Omega ^2 x^2\big)
\Big]^2\cr
&&+\mathbb{L}\Big(\frac{dr^2}{Q}+\frac{dx^2}{(1-x^2)P}\Big)\,,\label{Rotating Ricci flat}
\eea
where the two functions (\(\mathbb{M}\), \(\mathbb{L}\)) are somewhat complicated, given explicitly by
\bea
\mathbb{M}&=&\frac{1}{2 I_1^3 \Omega }\bigg\{
\Omega  \Big[I_1 I_2 \mu  r x^2 \big(B^2 \mu  r (B^2 r^2+x^2)+2 I_1 (B^2 r^2+1)\big)-I_1^3 \Sigma  (B^2 r^2 x^2+1)\Big]\cr
&&+(B^2 r^2+1)\Big[
I_1^3 \Sigma  (a^2 B^2 x^2-1)+I_2 \mu  r x^2 \big(I_2 B^4 \mu ^2 r^2 x^2+3 I_1 I_2 B^2 \mu  r x^2\cr
&&-I_1^2 (3 a^2 B^2 x^2+B^2 r^2 x^2-2)\big)
\Big]
\bigg\}\,,\cr
\mathbb{L}&=&\frac{1}{4 I_1^2 \Omega ^4}\Big\{
\Sigma  \big[B^4 r^2 \mu ^2 I_2 x^2+I_1^2 \big(2+a^2 B^4 r^2 x^2+B^2 (r^2-a^2 x^2-r^2 x^2)\big)\big]\cr
&&+B^2 r \mu  I_2 x^2 \big(2 a^2 I_1
   x^2+B^2 r^3 \mu  I_2 x^2+2 r^2 I_1 (2-a^2 B^2 x^2)\big)\cr
   &&+2 \Omega  I_1 (\Sigma  I_1+B^2 r^3 \mu  I_2 x^2)
\Big\}\,.
\eea
The remaining quantities follow the conventions of \cite{Podolsky:2025tle, Ovcharenko:2025cpm} and are given by
\bea
\Sigma&=&r^2+a^2x^2\,,\qquad
P=1+B^2\Big(\mu^2\frac{I_2}{I_1^2}-a^2\Big)x^2\,,\qquad Q=(1+B^2r^2)\Delta\,,\cr
\Omega^2&=&(1+B^2r^2)-B^2 x^2 \Delta\,,\qquad \Delta=\Big(1-B^2\mu^2\frac{I_2}{I_1^2}\Big)r^2-2\mu\frac{I_2}{I_1}r+a^2\,,\cr
I_1&=&1-\frac{1}{2}B^2a^2\,,\qquad I_2=1-B^2a^2\,.\label{4D solution 2}
\eea
The metric is cohomogeneity-two, depending on the radial $r$ and latitude angular $x=\cos\theta$ coordinates. There are three nontrivial parameters $(B,\mu, a)$. The parameter $P_0$ is fixed by requiring the azimuthal coordinate $\phi$ to have a period of $2\pi$. The general solution is of Petrov type I.

\subsection{Special limits}

When constant $B$ vanishes, \eqref{Rotating Ricci flat} reduces to the Kerr metric in standard Boyer-Lindquist form. In the limit of $a\rightarrow 0$, the metric becomes static and is given by
\be
ds^2=\frac{(1+\Omega+B^2 \mu  r x^2)^2}{4\Omega^4}\Big[-\frac{Q}{r^2}dt^2+\frac{r^2}{Q}dr^2+\frac{r^2 dx^2}{(1-x^2)P}\Big]
+\frac{4r^2(1-x^2)P}{P_0^2(1+\Omega+B^2 \mu  r x^2)^2}d\phi^2\,.\label{4D Ricci-flat}
\ee
This static metric was previously reported and analysed in \cite{Astorino:2026okd}. Expectedly, it becomes the Schwarzschild metric of mass $\mu$ when we take $B=0$. When both $a$ and $\mu$ vanish, the spacetime becomes a warped product of $U(1)$ and AdS$_3$, with characteristic length scale $\ell=1/B$. The Ricci-flat metric is
\bea
ds^2 &=& \frac{\left(1+\sqrt{1+ B^2 r^2 \sin ^2\theta}\right)^2 }{4 \left(1 +B^2 r^2 \sin ^2\theta\right)^2} ds_{{\rm AdS}_3}^2+\frac{4 r^2 \sin^2\theta}{\left(1+\sqrt{1+B^2 r^2 \sin ^2\theta}\right)^2} d\phi^2\,,\nn\\
ds_{{\rm AdS}_3}^2 &=&\frac{dr^2}{1+B^2 r^2}- \left(1+ B^2 r^2\right) dt^2+ r^2 d\theta^2\,.
\eea
As was observed in \cite{Astorino:2026okd}, this metric is free of singularities and is of Petrov type D within the Kundts class of solutions, with Riem$^2= 768 B^4 (B^2 r^2 \sin ^2\theta +1)^3/(\sqrt{B^2 r^2 \sin ^2\theta+1}+1)^6$. There is no asymptotically-flat region for real $(r,\theta)$ coordinates. We find that the radial direction is compact, except in the north and south poles. This can be seen from the total ``comoving'' radial distance for fixed angle $\theta$, given by
\be
L_r(\theta) \equiv \int_{0}^\infty \sqrt{g_{rr}}\, dr = \ft{1}{2B} \left(K(\cos ^2\theta )+\sec\theta \log \cot(\ft12\theta)\right),
\ee
where $K(x)$ is the elliptic $K$ function. At the equator, $L_r(\pi/2)=(2+\pi)/(4B)$, and it exhibits a logarithmic divergence at both the north and south poles. The $r\in [0,\infty)$ region of the spacetime is thus the interior of a regular surrounding dome of a spindle shape, with two infinite but shrinking openings at north and south poles.  We refer to this region of the spacetime as ``B-spindle.'' The spindle geometry implies that the spacetime described by the $(r, x)$ coordinates is not geometrically complete.
See \cite{Astorino:2026okd} for further analysis.

\section{Global analysis and black hole thermodynamics}

We have shown that our new rotating metric in the $(r,x)$ coordinates describes a spacetime inside the B-spindle. However, turning on $(\mu,a)$ alters the asymptotic structure of the B-spindle, and we therefore refer to the resulting geometry as the deformed B-spindle. The major difference is that the two asymptotically-infinite openings at the poles now close off, resulting in a finite spacetime volume. The metric possesses four degenerate hypersurfaces. Two are located at the north and south poles, corresponding to $x=\mp 1$. To ensure that the metric is free of a closed timelike curve (CTC), it is necessary to perform a linear coordinate transformation,
\be
t\quad \rightarrow\quad t'=t-\frac{2a}{(1+\sqrt{I_2})P_0} \phi\,,\label{remove the CTCs}
\ee
so that the null Killing vector at $x=\pm 1$ is simply \(\ell_{\phi} = \partial_{\phi}\), with vanishing time component. Imposing the periodicity condition $\Delta\phi=2\pi$, we obtain
\be
P_0=\frac{2I_2(I_1^2+B^2\mu^2)}{I_1^2(1+\sqrt{I_2})}\,.\label{P0cons}
\ee
This ensures that $\phi$ is the only periodic Killing coordinate. However, CTCs can still exist when $g_{\phi\phi}$ becomes negative. For our metric obtained after the coordinate transformation \eqref{remove the CTCs}, we have
\bea
g_{\phi\phi}&=&\frac{I_1^4}{4I_2^2(I_1^2+B^2\mu^2)^2}\,\frac{1}{r^4(1-x^2)P-a^2x^4Q}
\Bigg\{-4 a^2(1-x^2) P Q  \mathbb{L}\cr
&&+ \frac{1}{\mathbb{L} \Omega ^4}\Big[(1+\sqrt{I_2}) \big(P r^4 (1-x^2)-a^2 Q x^4\big)-2 a^2 \mathbb{M}\Big]^2\Bigg\}.\label{gvpvp}
\eea
In the static $a=0$ limit, $g_{\phi\phi}$ is positive except at $x=\pm 1$, where it vanishes, as discussed above. In rotating spacetimes, however, CTCs may occur, as in the case of an over-rotating Kerr black hole. A physically acceptable black hole should have any CTCs confined within the event horizon, if they exist. We thus examine the conditions for the existence of the event horizon next and return to the issue of CTCs later.

The remaining two degenerate surfaces are null, located at the two positive roots $r_+>r_-$ of the function $Q$, determined by the quadratic function $\Delta(r_\pm)=0$. In order to achieve this, we must impose that the parameter $\mu$ must satisfy
\be
\mu<\mu_{\max} \equiv \fft{I_1}{B\sqrt{I_2}}\,,
\ee
so that the coefficient of $r^2$ of $\Delta(r)$ is positive. This requirement ensures that $r$ is a non-compact coordinate extending to $r\rightarrow \infty$ since $\Delta(r)$ is positive for large $r$.  The two positive roots of the quadratic function $\Delta(r)$ are then given by
\be
r_\pm = \frac{I_1 (I_2 \mu \pm \sqrt{I_2\mu^2 - a^2 I_1^2})}{B^2I_2(\mu_{\rm max}^2-\mu^2)}\,,
\ee
provided that $\mu$ satisfies
\be
\mu_{\rm max}>\mu \ge \mu_{\rm ext} \equiv \fft{a I_1}{\sqrt{I_2}}\,,\qquad\rightarrow\qquad aB<1\,.
\ee
When all these conditions are satisfied, $\Delta(r)=0$ generally has two positive roots, with the larger root $r_+$ corresponding to the horizon radius.

Since the curvature singularity lies inside $r_-$ and the spacetime is regular from $r_+$ to asymptotic infinity, the outer horizon $r_+$ can be viewed as the event horizon. Since the asymptotic geometry is the deformed B-spindle rather than asymptotically flat, there is no canonical normalization of the timelike Killing vector. The absence of naked CTCs and the requirement $\Delta \phi=2\pi$ imply that we can introduce the following most general linear transformation
\be
t\,\,\rightarrow\,\, t' = \lambda_1\, t\,,\qquad \phi\,\,\rightarrow\,\, \phi'=\phi + \lambda_2 B\, t\,,\label{lambda12}
\ee
where $(\lambda_1,\lambda_2)$ are two dimensionless constants to be determined. It is important to note that the above transformations do not affect $g_{\phi\phi}$ and hence the discussion of CTCs.

{\bf Komar conserved quantities}: Although there is no well-defined notion of mass,
the spacetime admits two commuting Killing vectors $\xi=c_1 \partial_t - c_2 \partial_\phi$, giving rise to two conserved charges. The Komar 2-form, defined by $\mathbf{Q} = - {*d\xi}$, is close for our Ricci-flat metric, i.e.~$d\mathbf{Q}=0$. This implies that \(\mathbf{Q}\) takes form
\be
\mathbf{Q}=V(r,x) d\phi\wedge dx+U(r,x)dr\wedge d\phi\,,\qquad \hbox{with}\qquad \partial_rV+\partial_{x}U=0\,.
\ee
According to \cite{Liu:2022wku}, we then have constant integrations
\bea
c_1 M + 2 c_2 J&=&\frac{1}{8\pi}\int \mathbf{Q}=\frac{\int d\phi}{8\pi}\Big(
\int_{-1}^{+1}V(r,x)dx+\int_{r_h}^rU(r',x)\Big|_{x=-1}^{x=+1}dr'
\Big)\,.
\eea
Here, $(c_1,c_2)$ are two auxiliary parameters used to extract the conserved charges $(M,J)$, which are precisely the mass and angular momentum of the Kerr back hole when $B=0$. For $B\ne 0$, $J$ is independent of $(\lambda_1,\lambda_2)$ and can still be interpreted as the angular momentum. Whether $M$ can be identified as the (thermodynamic) mass depends on whether the first law of black hole thermodynamics can be established.

{\bf Horizon geometry and data}: As discussed above, the horizon is located at $r=r_+$, the largest positive root of quadratic $\Delta(r)$. We thus have
\be
\mu=\frac{\left(\alpha ^2+1\right) (\beta -\alpha )}{2 \alpha  \sqrt{\beta ^2-1} B}\,,\qquad
\alpha = \sqrt{1-B^2 a^2}\,,\qquad \beta = \sqrt{1 + B^2 r_+^2}\,.
\ee
For notational convenience, we have introduced two positive dimensionless parameters $\beta \ge 1\ge \alpha$. To make the $\mathbb{R}^2\times S^2$ structure of the near-horizon geometry manifest, we can rewrite the metric in the following vielbein basis:
\be
ds^2=\Big(-a_1dt^2 + g_{rr}dr^2\Big) + \Big(a_2(d\phi+a_3dt)^2+g_{xx}dx^2\Big)\,,
\ee
where $a_1=-g_{tt}+g_{t\phi}^2/g_{\phi\phi}$,  $a_2=g_{\phi\phi}$ and $a_3=g_{t\phi}/g_{\phi\phi}$. As $r\rightarrow r_+$, the two brackets become $\mathbb{R}^2$ and $S^2$ respectively. The entropy is given by $S=1/4 \int \sqrt{a_2 g_{xx}} dx d\phi$.
Similar to the angular momentum, the entropy is independent of $(\lambda_1,\lambda_2)$. The temperature and angular velocity are
\be
T=\frac{1}{4\pi}\sqrt{\frac{a_1}{g_{rr}}}\frac{Q'}{Q}\Big|_{r=r_+}\,,\qquad
\omega=a_3(r_+)\,.
\ee
We have now computed all five would-be thermodynamic quantities: $(M,T,\omega)$ depend on $(B,\mu,a, \lambda_1,\lambda_2)$, while $(J,S)$ depend only on $(B,\mu,a)$. When $B=0$ and upon setting $\lambda_1=1$, these quantities reduce to those of the Kerr black hole, which satisfy the first law and the Smarr relation
\be
dM=TdS+ \omega dJ\,,\qquad M=2 TS + 2\omega J\,.\label{flsmar}
\ee
In the above canonical ensemble, the Kerr black hole mass is a function of its entropy and angular momentum, explicitly given by
\be
M^2 = \fft{S}{4\pi} + \fft{\pi J^2}{S}\,.\label{MSJ}
\ee
Now for $B\ne 0$, we find that the Smarr relation still holds, and is independent of $(\lambda_1,\lambda_2)$. We are unable to determine parameter $(\lambda_1, \lambda_2)$ from the first law alone. We therefore impose \eqref{MSJ} even when $B\ne 0$. Under this condition, we find that $(\lambda_1,\lambda_2)$ must be
\bea
\lambda_1 &=& \frac{2 \alpha  (\alpha +1) \left(\beta ^2-1\right)-\gamma ^2}{\gamma \sqrt{\beta (\alpha +1)(\alpha ^2 \beta -2 \alpha +\beta)}}\,,\nn\\
\lambda_2 &=& -\frac{\sqrt{\beta (1-\alpha )\left(\alpha ^2 \beta -2 \alpha +\beta \right)} \left((\alpha +1)^2 \beta ^2-2 \alpha  \beta -2 \alpha \right)}{(\alpha +1) \left(\beta ^2-1\right) \gamma }\,,
\eea
where $\gamma = \sqrt{\left(1+\alpha + \alpha^2 -\alpha ^3\right) \beta ^2+2 (\alpha -1) \alpha  \beta -2 \alpha }$. Now, all the thermodynamic quantities are fully determined, given by
\bea
M &= &\frac{\gamma  (\beta -\alpha )  \sqrt{(\alpha +1)(\beta^2-1)}}{2 \beta ^{3/2} \left(\alpha ^2 \beta -2 \alpha +\beta \right)^{3/2}B}\,,\qquad
J = \frac{(\beta -\alpha ) (\alpha +1)^2 \sqrt{1-\alpha ^2} \left(\beta ^2-1\right)^{3/2} }{4 \beta ^2 \left(\alpha ^2 \beta -2 \alpha +\beta \right)^2 B^2}\,,\cr
T &=& \frac{B\,\beta ^{3/2} (\alpha  \beta -1) \left(2 \alpha  (\alpha +1) \left(\beta ^2-1\right)-\gamma ^2\right)}{2 \pi  \gamma (\beta -\alpha )(\beta +1) \sqrt{(\alpha +1)(\beta ^2-1)(\alpha ^2 \beta -2 \alpha +\beta)}}\,,\\
S&=&\frac{\pi  (\alpha +1) (\beta -1) (\beta +1)^2 (\beta -\alpha )}{2 \beta ^3 \left(\alpha ^2 \beta -2 \alpha +\beta \right)B^2}\,,\quad
\omega = \frac{B (\alpha +1) \beta ^{5/2} \sqrt{(1-\alpha)(\alpha ^2 \beta -2 \alpha +\beta)}}{\gamma (\beta -\alpha ) (\beta +1)}\,.\nn
\eea
It can be easily checked that these thermodynamic quantities satisfy both \eqref{flsmar} and \eqref{MSJ}. The spindle parameter $B$ does not introduce a new thermodynamic variable. In the extremal zero-temperature limit, corresponding to the condition $\alpha\beta=1$, there exists a decoupling limit that lead to a $U(1)$ squashed AdS$_3$ factor in the near-horizon geometry. Interestingly, the Kerr relation $S_{\rm ext}=2\pi J$ in the extremal limit still holds in the more general spindle case. The imposition of \eqref{MSJ} is consistent with the thermodynamic framework of \cite{Christodoulou:1971pcn}; however, we established the first law at the metric level by constructing an explicit coordinate transformation \eqref{lambda12}. Although the thermodynamics of the type-D KBR black hole \cite{Podolsky:2025tle,Ovcharenko:2025cpm} was already analysed in \cite{Hu:2026slp}, our analysis of the Ricci-flat black hole may shed new light on the subtleties in the KBR thermodynamics.

Since we need to perform the coordinate transformation \eqref{lambda12} and introduce specifically-defined $(\lambda_1,\lambda_2)$ parameters to establish the thermodynamic quantities that satisfy both the first law and CR relation \eqref{MSJ}, we need examine whether the metric in new coordinates is still well defined. Specifically, we examine the CTCs and ergoregions. As discussed above, the metric component $g_{\phi\phi}$ in \eqref{gvpvp} is independent of $(\lambda_1,\lambda_2)$. It can be readily verified that, for $\mu\in (\mu_{\rm max}, \mu_{\rm ext}]$, $g_{\phi\phi}$ is a monotonically decreasing function of $x^2 \in [0,1]$, vanishing at $x^2=1$. At $x=0$, $g_{\phi\phi}$ decreases from infinity as $r\rightarrow 0$ to a positive minimum inside the horizon, and then increases to a finite positive quantity at asymptotic infinity. Specifically, we have
\bea
g_{\phi\phi} (r=r_+,x=0) &=& \frac{1}{4(1+ B^2 r_+^2)}
\bigg(\frac{r_+(1+\sqrt{I_2}\,) \big(\sqrt{I_2} - \sqrt{1 + B^2 r_+^2}\,\big)}{
I_1 \sqrt{1 + B^2 r_+^2} - \sqrt{I_2}}\bigg)^2\,,\nn\\
g_{\phi\phi} (r\rightarrow\infty,x=0) &=& \frac{I_1^4 (1 + \sqrt{I_2}) (3 + 5\sqrt{I_2}) + a^2 B^4 I_1^2 I_2 \mu^2}{4B^2 I_2^2 (I_1^2 + B^2 \mu^2)^2}>0\,.
\eea
Thus, the $B$-deformed Kerr metric is free of CTCs, just like the (undeformed) Kerr black hole.

We now examine the ergoregions of our $B$-deformed Kerr black hole. The ergoregion is defined by $g_{tt}>0$, and its existence is inevitable in rotating black hole spacetimes.
The issue is whether such an ergoregion extends all the way to infinity. Indeed, the ergoregion of a Kerr-Melvin black hole extends to infinity at vicinities of the north and south poles. One salient feature of the KBR metric is that its finite ergoregion is finite \cite{Podolsky:2025tle}. However, this issue becomes more subtle when we impose the first law, which requires a coordinate transformation of the Killing coordinates, such as \eqref{lambda12}. The parameter $\lambda_2$ specifies the choice of comoving coordinate and thus inevitably affects the metric component $g_{tt}$. We therefore consider two situations. The first is the metric obtained after coordinate transformation \eqref{remove the CTCs}, under which $\phi$ is the only periodic Killing coordinate. We denote the time component of the metric by $\tilde g_{tt}$. The second is the metric obtained after the additional transformation \eqref{lambda12}, which is required to satisfy the first law. We denote the corresponding time component of the metric as $\hat g_{tt}$.

We first examine $\tilde g_{tt}$. At asymptotic infinity, $\tilde g_{tt}$ attains its maximum at $x=0$, where it is given by
\be
\tilde g_{tt} (r\rightarrow \infty, x=0) = \frac{B^2 I_2}{4 I_1^2} (\mu^2 - \mu_{\rm max}^2)<0\,.
\ee
Thus, the ergoregions are all finite in this case. The situation is more complicated for the $\hat g_{tt}$. Nevertheless, we find that at the asymptotic infinity, $\hat g_{tt}$ also attains its maximum at $x=0$, where it is given by
\be
E=\hat g_{tt}(r\rightarrow \infty,x=0) =\frac{4}{P_0^2}
\bigg(\lambda_2^2 - \frac{B^2I_2\mu^2-I_1^2}{16I_1^2} \Big(P_0\lambda_1 -
\frac{2aB\lambda_2}{1+\sqrt{I_2}}\Big)^2\bigg).
\ee
In this case, the ergoregions are not always finite. However, for suitable parameters, black holes with finite ergoregions do exist. As an example, we consider $(B,a)=(1/10,1)$, for which we obtain
\be
\mu_{\rm max}=\frac{199}{6\sqrt{11}}\sim 10.00\,,\qquad \mu_{\rm ext}=\frac{199}{60\sqrt{11}}\sim 1.000\,.
\ee
We find that $\hat g_{tt}$ is negative throughout the asymptotic $r\rightarrow \infty$, provided that the parameter $\mu$ lies in
\be
\mu_{\rm ext} \le \mu < 9.22 <\mu_{\rm max}\,.
\ee
We now provide a concrete example with $B=1/10$, $a=1$ and $\mu=5$. In this case, the inner and outer horizons are
\be
r_-=0.1013\,,\qquad r_+=13.16\,.
\ee
We find that the ergoregion is confined to $r_+<r<r_*$, where $r_*=13.88$, which is very close to $r_+$. By comparison, for the Kerr black hole with the same $(\mu,a)=(5,1)$, the horizon radii are $r_+=9.899$ and $r_-=0.1010$, and outer boundary of the ergoregion is at $r_*=10$. The thermodynamic quantities of the $B$-deformed black hole are
\be
M=3.613\,,\qquad J=3.236\,,\qquad S=161.5\,,\qquad T=0.01084\,,\qquad \omega=0.01742\,.
\ee
For the Kerr black hole with $\mu=5$ and $a=1$, the thermodynamic quantities are
\be
M=5\,,\qquad J=5\,,\qquad S=311.0\,,\qquad T=0.007877\,,\qquad \omega=0.01010\,.
\ee

\section{Re-magnetizing the Ricci-flat metric}

We can re-magnetize the Ricci-flat metric in a presence of an independent uniform magnetic field ${\cal B}$. As an illustrative example, we consider the magnetization of the static solution \eqref{4D Ricci-flat}, which reproduces a special case of the static solutions obtained in \cite{Astorino:2025lih}:
\bea
ds^2&=&\frac{1}{\Omega^2}\Bigg[\bigg\{
1-\frac{r^2(B^2 P_0^2-\mathcal{B}^2) (1-x^2)P }{4 \Omega ^2 P_0^4 (1+\Omega +B^2 r \mu  x^2)^2} \Big[r^2 \mathcal{B}^2 (1-x^2) P\cr
&&+P_0^2 (1+\Omega +B^2
   r \mu  x^2) (1+3 \Omega +B^2 r \mu  x^2)\Big]
\bigg\}\Big[-\frac{Q}{r^2}dt^2+\frac{r^2}{Q}dr^2+\frac{r^2 dx^2}{(1-x^2)P}\Big]\cr
&&+\bigg\{
1+\frac{r^2 (B^2 P_0^2-\mathcal{B}^2) (1-x^2) P}{\big[r^2 \mathcal{B}^2 (1-x^2) P+P_0^2 (1+\Omega +B^2
   r \mu  x^2)^2\big]^2}\Big[r^2 \mathcal{B}^2 (1-x^2)P\cr
   &&+P_0^2 (1+\Omega +B^2 r \mu  x^2) (1+3 \Omega +B^2 r \mu  x^2)
   \Big]
\bigg\}\fft{r^2 (1-x^2)P}{P_0^2} d\phi^2\Bigg]\,,\cr
A_{\1}&=&\fft{2{\cal B} \Omega\,  (1 - \Omega + r \partial_r\Omega)}
{(B^2 P_0^2 - {\cal B}^2) (1 + B^2\mu r x^2 \Omega) + 2 {\cal B}^2 \Omega}\,d\phi\,.
\label{staticKBR}
\eea
This solution contains both the spindle parameter $B$ and the external magnetic field ${\cal B}$. When $B=0$, the solution becomes the Schwarzschild-Melvin solution. When \(\mathcal{B} = \pm B P_0\), it reduces to the static magnetized solution of \cite{Podolsky:2025tle,Ovcharenko:2025cpm}. Therefore, the solution \eqref{staticKBR} with independent $(B, {\cal B})$ provides a unified description of the type-I Melvin and the static type-D KBR solutions. Importantly, the periodicity condition $\Delta\phi=2\pi$ is independent of the value of the external magnetic field ${\cal B}$.

The magnetization of the more general rotating case were discussed in the appendix. Additional subtleties arise when rotation is involved. A natural choice for the KK reduction would be on the $\phi$, after removing naked CTCs using \eqref{remove the CTCs}. However, in order to recover the magnetized rotating KBR solution, the KK reduction must be performed along the original $\phi$ direction of the metric \eqref{Rotating Ricci flat}. Conversely, we must first make the coordinate transformation $t\rightarrow t + a \phi/P_0$ on the original KBR solution before carrying out the KK reduction in order to recover the Ricci-flat metric \eqref{Rotating Ricci flat}. Consequently, the requirement of $\Delta\phi=2\pi$ leads to a ${\cal B}$-dependent $P_0$. This suggests that our new rotating metric serves as a neutral seed for constructing a variety of inequivalent magnetization schemes of the Kerr black hole.

\section{Conclusion}

In this paper, we have obtained a new Ricci-flat metric by demagnetizing the recently constructed KBR solution. The metric contains three independent parameters, $(B,\mu,a)$. The parameter $B$, which corresponds to the uniform magnetic field in KBR solution, is therefore intrinsically geometric. When $B$ vanishes, it reduces to the standard Kerr metric. Turning on $B$ has the effect that the original non-compact radial coordinate becomes compact so that $r\rightarrow \infty$ describes a regular dome of spindle shape with shrinking finite or infinite north and south poles, depending on the parameters. The spindle geometry indicates that the spacetime described by the $(r,x)$ coordinates is not geodesically complete.

Despite lacking an asymptotically-flat region, we find that the first law of black hole thermodynamics can be recovered upon a suitable rescaling of time and azimuthal coordinates. The thermodynamic relations \eqref{flsmar} and \eqref{MSJ} are identical to those of the Kerr black hole, as if the constant $B$ were absent. Remarkably, the extremal Kerr relation $S_{\rm ext} =2\pi J$ still holds, independent of $B$.

We re-magnetize the Ricci-flat solution in the presence of a uniform external magnetic field ${\cal B}$, and recover the KBR solution only after identifying ${\cal B}$ with the spindle parameter $B$. Thus, the magnetic field parameter in the KBR solution is fundamentally different from that in the Melvin solutions, where the demagnetization transformation is equivalent to setting $B=0$. Our Ricci-flat rotating solution therefore provides a neutral seed for a variety of inequivalent magnetizations of the Kerr black hole.

\section*{Acknowledgement}

We are grateful to Zhi-Chao Li, Pujian Mao and Yi Pang for useful discussions. L.M.~is supported in part by National Natural Science Foundation of China (NSFC) grant No.~12447138, Postdoctoral Fellowship Program of CPSF Grant No.~GZC20241211, the China Postdoctoral Science Foundation under Grant No.~2024M762338 and the National Key Research and Development Program No.~2022YFE0134300. H.L.~is supported in part by the NSFC grants No.~12375052 and No.~11935009. The work is also supported in part by the Tianjin University Self-Innovation Fund Extreme Basic Research Project Grant No.~2025XJ21-0007.

\appendix

\section{(De)magnetization in Einstein-Maxwell theory}

In this appendix, we first discuss the (de)magnetization transformation in EM gravity
\bea
\mathcal{L}=\sqrt{-g}(R-F^2)\,,\qquad F^2=F^{\mu\nu} F_{\mu\nu}\,,\qquad F_{\mu\nu}=\partial_\mu A_\nu-\partial_\nu A_\mu\,.\label{4D EM}
\eea
We then magnetize the Ricci-flat rotating solution \eqref{Rotating Ricci flat} and verify that, for a suitable fine-tuning of the parameter, the resulting solution matches the magnetized KBR solution \cite{Podolsky:2025tle}.

\subsection{A brief review of global $SU(2,1)$ symmetry}

We follow the procedure in \cite{Gibbons:2013yq} and review the magnetization transformation in EM gravity, while correcting some typos on the way. We begin by performing the \(S^1\) KK reduction from four to three dimensions:
\be
ds^2=e^{2\varphi}d\bar{s}_3^2+e^{-2\varphi}(dz+2\bar{\mathcal{A}}_{\1})^2\,,\qquad
A_{\1}=\bar{A}_{\1}+\chi(dz+2\bar{\mathcal{A}}_{\1})\,.\label{reduction ansatz}
\ee
The resulting three-dimensional theory is
\bea
\bar{\mathcal{L}}_3&=&\sqrt{-\bar g}\Big(\bar{R}-2(\partial\varphi)^2-2e^{2\varphi}(\partial\chi)^2
-e^{-4\varphi}\bar{\mathcal{F}}^2-e^{-2\varphi}\bar{F}^2\Big)\,,\cr
\bar{\mathcal{F}}_{\2}&=&d\bar{\mathcal{A}}_{\1}\,,\qquad \bar{F}_{\2}=d\bar{A}_{\1}+2\chi d\bar{\mathcal{A}}_{\1}\,.\label{3D action F}
\eea
We then adopt the first-order formalism, introducing two Lagrange multipliers $(\psi,\sigma)$ to dualize the two 2-form field strengths to become scalar fields. For simplicity, we express the Lagrangian in differentia-form notation
\bea
\bar{\mathcal{L}}_3&=&\bar{R}\,{*\oneone}-2{*d}\varphi\wedge d\varphi-2e^{2\varphi}{*d}\chi\wedge d\chi-2e^{-4\varphi}{*\bar{\mathcal{F}}_{\2}}\wedge\bar{\mathcal{F}}_{\2}-2e^{-2\varphi}{*\bar{F}_{\2}}\wedge \bar{F}_{\2}\cr
&&+ 4d\psi\wedge(\bar{F}_{\2}-2\chi\bar{\mathcal{F}}_{\2})+ 4d\sigma\wedge\bar{\mathcal{F}}_{\2}\,.\label{3D action F scalar}
\eea
The equations of motion for the scalar fields $(\psi,\sigma)$ give rise to two Bianchi identities
\bea
d\bar{\mathcal{F}}_{\2}=0\,,\qquad d(\bar{F}_{\2}-2\chi\bar{\mathcal{F}}_{\2})=0\,,
\eea
which recover the expressions for the field strengths in the second-order formalism \eqref{3D action F}. By substituting the equations of motion for \(\bar{F}_{\2}\) and \(\bar{\mathcal{F}}_{\2}\),
\bea
-4e^{-4\varphi}{*\bar{\mathcal{F}}_{\2}}- 8\chi d\psi+ 4d\sigma=0\,,\qquad
-4e^{-2\varphi}{*\bar{F}_{\2}}+4d\psi=0\,,\label{F dual to scalar}
\eea
back into \eqref{3D action F scalar}, we obtain the corresponding dual three-dimensional Einstein-scalar theory
\bea
\bar{\mathcal{L}}_3&=&\bar{R}\,{*\oneone}-2{*d}\varphi\wedge d\varphi-2e^{2\varphi}{*d}\chi\wedge d\chi-2e^{2\varphi}{*d}\psi\wedge d\psi\cr
&&-2e^{4\varphi}(*d\sigma-2\chi {*d}\psi)\wedge(d\sigma-2\chi d\psi)
\eea
or
\be
\bar{\mathcal{L}}_3=\sqrt{-\bar g}\Big(\bar{R}-2(\partial\varphi)^2
-2e^{2\varphi}(\partial\chi)^2-2e^{2\varphi}(\partial\psi)^2
-2e^{4\varphi}(\partial\sigma-2\chi \partial\psi)^2\Big)\,.
\label{3D action scalar}
\ee
The scalar sector constitutes a coset of \(SU(2,1)/(SU(2)\times U(1))\). Consequently, any solution of \eqref{3D action scalar} is mapped to aother solution under a global nonlinear \(SU(2,1)\) symmetry transformation.

To find an explicit (de)magnetization transformation, we define a $3 \times 3$ matrix $E_a^{{\color{white}a}b}$ for $a,b = 0,1,2$, in which all entries are zero except for the element at the $a$-th row and $b$-th column, which is set to 1. We can parameterise a coset representative as
\bea
\mathcal{V}=e^{\varphi H}e^{-2i\sigma E_0^{{\color{white}2}2}}e^{\sqrt{2}\chi( E_0^{{\color{white}2}1}+E_1^{{\color{white}2}2})}
e^{-i\sqrt{2}\psi( E_0^{{\color{white}2}1}-E_1^{{\color{white}2}2})}\,,\qquad H=E_0^{{\color{white}2}0}-E_2^{{\color{white}2}2}\,.
\eea
One can verify that $\mathcal{V}$ belongs to $SU(2,1)$, which can be confirmed using the invariant metric $\eta$ of $SU(2,1)$
\be
\mathcal{V}^\dag\eta \mathcal{V}=\eta\,,\qquad \eta=\begin{pmatrix}
0 & 0 & -1 \\
0 & 1 & 0 \\
-1 & 0 & 0
\end{pmatrix}\,.
\ee
Defining $\mathcal{M}=\mathcal{V}^\dag \mathcal{V}$,  the Lagrangian \eqref{3D action scalar} becomes $\bar{\mathcal{L}}_3=\sqrt{-\bar g}\Big(\bar{R}
-\frac{1}{4}[\mathrm{tr}(\mathcal{M}^{-1}\partial\mathcal{M})]^2\Big)$.
For any constant $SU(2,1)$ transformation matrix $U$ with $U^\dagger \eta U = \eta$, the Lagrangian is invariant under the associated $SU(2,1)$ transformations
\be
\mathcal{M}\rightarrow\widetilde{\mathcal{M}}=U^\dagger \mathcal{M} U\,.\label{constant SU21}
\ee
The (de)magnetization transformation matrix
\be
U =
\renewcommand{\arraystretch}{1.5}
\begin{pmatrix}
1 & 0 &\; 0 \\
\displaystyle \frac{\mathcal{B}}{\sqrt{2}} & 1 &\; 0 \\
\displaystyle \frac{\mathcal{B}^2}{4} & \displaystyle \frac{\mathcal{B}}{\sqrt{2}} &\; 1
\end{pmatrix}\,\label{Magnetizing U}
\ee
is then a specific constant matrix belonging to \(SU(2,1)\). For example, the Schwarzschild black hole maps to the Melvin universe under a uniform external magnetic field B.

As a simple illustration, we may begin with the Ricci-flat static solution \eqref{4D Ricci-flat}, and apply the magnetizing transformation to otain \eqref{staticKBR}. The resulting solution shows that the $B$-spindle parameter is generally distinct from the external constant magnetic field ${\cal B}$.

\subsection{Magnetizing the new Ricci-flat rotating solution}

We now re-magnetize the new general rotating metric \eqref{Rotating Ricci flat} following the same transformation. In order to reproduce the result of \cite{Podolsky:2025tle, Ovcharenko:2025cpm}, it is necessary to perform the KK reduction along the original $\phi$ direction in \eqref{Rotating Ricci flat}, which involves naked CTCs. The resulting three-dimensional solution is
\bea
d\bar{s}_3^2&=&-\frac{PQ}{P_0^2\Omega^4}(1-x^2)dt^2+\frac{
P r^4 (1-x^2)-a^2 Q x^4}{P_0^2\Omega^4}\Big(\frac{1}{P}\frac{dx^2}{1-x^2}+\frac{dr^2}{Q}\Big)\,,\cr
\bar{\mathcal{A}}_{\1} &=& \frac{a P_0 \mathbb{M}}{2 \big[r^4 (1-x^2)P-a^2 x^4 Q\big]}dt\,,\ \
e^{2\varphi}=\frac{ \Omega ^4 P_0^2 \mathbb{L}}{r^4 (1-x^2)P-a^2x^4 Q}\,,\ \
\bar{A}_{\1}=0\,,\ \  \chi=0\,.
\eea
Using \eqref{F dual to scalar}, we can dualize the two 1-form fields\(\{\bar{A}_{\1},\bar{\mathcal{A}}_{\1}\}\) into two scalar fields \(\{\psi,\sigma\}\)
\bea
\psi=0\,,\qquad \sigma=\frac{a r x (\Sigma  I_1-r \mu  I_2 x^2)}{\Omega ^3 I_1 P_0^2 \mathbb{L}}\,,
\eea
while the other fields remain unchanged.

Next, we magnetize the four scalar fields by applying the \(SU(2,1)\) transformation \eqref{constant SU21}, with a specific magnetic field \(\mathcal{B} = -B P_0\) selected in the magnetization matrix \eqref{Magnetizing U}, following the static example. The transformed four scalar fields are
\bea
\tilde{\psi}&=&\frac{a B r x (r \mu  I_2 x^2-\Sigma  I_1)}{\Sigma  \Omega  I_1 P_0}\,,\qquad
\tilde{\sigma}=\frac{a r x (r \mu  I_2 x^2-\Sigma  I_1) \big[\Sigma  (1-2 \Omega ) I_1+B^2 r^3 \mu  I_2
   x^2\big]}{\Sigma ^2 \Omega ^2 I_1^2 P_0^2}\,,\cr
\tilde{\varphi}&=&\frac{1}{2}\log\frac{\Sigma  \Omega ^2 P_0^2}{r^4 (1-x^2) P-a^2 x^4 Q}\,,\qquad
\tilde{\chi}=\frac{\Omega-1}{BP_0}-\frac{1}{\Sigma BP_0}\Big[
a^2x^3\partial_{x}\Omega+r^2\partial_r\Omega
\Big]\,.
\eea
We then recover the magnetized 1-form fields using \eqref{F dual to scalar}
\bea
\tilde{\bar{\mathcal{A}}}_{\1}&=&-\frac{a P_0 \big[r^2 (1-x^2) P+x^2 Q\big]}{2 \big[r^4 (1-x^2) P-a^2 x^4 Q\big]}dt\,,\cr
\tilde{\bar{A}}_{\1}&=&-\frac{adt}{B \big[ r^4 (1-x^2) P-a^2 x^4 Q\big]}\Big\{
(1-\Omega ) \big[x^2 Q+r^2 (1-x^2) P\big]\cr
&&+rx\big[
r(1-x^2)P\partial_{x}\Omega+xQ\partial_r\Omega
\big]
\Big\}\,.
\eea
Finally, by applying the reduction ansatz \eqref{reduction ansatz}, we obtain a rotating solution in the Einstein-Maxwell theory \eqref{4D EM}
\bea
d\tilde{s}^2&=&\frac{1}{\Omega^2}\Bigg\{
-\frac{Q}{\Sigma}\Big[dt+\frac{ax^2}{P_0}d\phi\Big]^2+
\frac{\Sigma}{Q}dr^2+\frac{\Sigma}{P}\frac{dx^2}{1-x^2}
+\frac{P}{\Sigma}(1-x^2)\Big[adt-\frac{r^2}{P_0}d\phi\Big]^2
\Bigg\}\,,\cr
\tilde{A}_{\1}&=&\frac{1}{B}\Bigg\{\frac{r\partial_r\Omega}{\Sigma}
\Big(adt-\frac{r^2}{P_0}d\phi\Big)
-\frac{ax\partial_{x}\Omega}{\Sigma}\Big(dt+\frac{ax^2}{P_0}d\phi\Big)
+\frac{\Omega-1}{P_0}d\phi
\Bigg\}\,.\label{4D solution 1}
\eea
It can be seen that after applying the coordinate transformation on \(t\)
\bea
dt\quad \rightarrow\quad dt'=dt-\frac{a}{P_0}d\phi\,,\label{remove the CTCs 2}
\eea
the above solution reproduces precisely the magnetized black hole found in \cite{Podolsky:2025tle, Ovcharenko:2025cpm}. One should not fix the constant $P_0$, since now in order to ensure that the azimuthal angle \(\phi\) has a period of \(2\pi\), the constant \(P_0\) must be taken as
\be
P_0=1+B^2\Big(\mu^2\frac{I_2}{I_1^2}-a^2\Big),
\ee
which differs from \eqref{P0cons} for the Ricci-flat solution. This may related to the fact that the coordinate used for the KK reduction relating the magnetized and Ricci-flat solutions involve naked CTCs. A more general magnetization scheme involving independent ${\cal B}$ and a general choice of $\phi$ is therefore required to clarify the whole procedure.

\end{document}